\documentclass[12pt]{article}
\pdfoutput=1
\usepackage{amsmath,amssymb,bm,graphicx}

\def\vslash{\rlap{\hspace{0.02cm}/}{v}}

\def\nbslash{\rlap{\hspace{0.02cm}/}{\bar n}}

\usepackage{bm}

\begin{document}

\begin{titlepage}

\begin{flushright}
WSU-HEP-1905\\
August 7, 2019\\
\end{flushright}

\vspace{0.7cm}
\begin{center}
\Large\bf\boldmath
Reevaluating Uncertainties in $\bar B\to X_s\gamma$ Decay
\unboldmath
\end{center}

\vspace{0.8cm}
\begin{center}
{\sc Ayesh Gunawardana$^{a}$ and Gil Paz$^{\, a,b,c}$}\\
\vspace{0.4cm}
{\it 
$^{a}$\! Department of Physics and Astronomy \\
Wayne State University, Detroit, Michigan 48201, USA\\}
\vspace{0.3cm}
{\it 
$^{b}$ Physics Department, Technion - Israel Institute of Technology,\\ 
Haifa 3200003, Israel\\}
\vspace{0.3cm}
{\it 
$^{c}$ Department of Particle Physics and Astrophysics, \\
Weizmann Institute of Science, Rehovot, Israel 7610001
}
\end{center}
\vspace{1.0cm}
\begin{abstract}
  \vspace{0.2cm}
  \noindent
The rare decay $\bar B\to X_s\gamma$ is an important probe of physics beyond the standard model. The largest uncertainty on the total rate and the CP asymmetry arises from resolved photon contributions.  These appear first at order $1/m_b$ and are related to operators other than $Q_{7\gamma}$ in the effective weak Hamiltonian. One of the three leading contributions,  $Q^q_1-Q_{7\gamma}$,  is described by a non-local function whose moments are related to HQET parameters. We use recent progress in our knowledge of these parameters to reevaluate  the resolved photon contribution to $\bar B\to X_s\gamma$  total rate and CP asymmetry.

\end{abstract}
\vfil

\end{titlepage}

\section{Introduction}
The rare decay $\bar B\to X_s\gamma$ is an important probe of physics beyond the standard model. Both the CP averaged branching ratio and the CP asymmetry are used to constrain many models of new physics. The latest theoretical prediction for the branching ratio within the Standard Model (SM) is $(3.36\pm0.23)\times10^{-4}$  \cite{Misiak:2015xwa} where a cut on the photon energy of $E_\gamma>1.6$ GeV is assumed. This can be compared to the 2019 update of the 2018 PDG (Particle Data Group) experimental value of $(3.49\pm0.19)\times10^{-4}$ \cite{Tanabashi:2018oca}. See also the HFLAV (Heavy Flavor Averaging Group) values in \cite{Amhis:2016xyh}. For the Belle II experiment the uncertainty is expected to be reduced \cite{Kou:2018nap}. This motivates an effort to reduce the uncertainty on the theoretical prediction.

The largest source of uncertainty in the theoretical prediction, $\sim5\%$, is non-perturbative effects in the form of resolved photon contributions. These arise  when the photon couples to light partons instead of coupling directly to the effective weak interaction. These effects were systematically studied in  \cite{Benzke:2010js}. They first appear at power $\Lambda_{\mbox{\scriptsize QCD}}/m_b$ and arise from the pairs of the weak hamiltonian operators $Q_1^q-Q_{7\gamma}$ (where $q=u,c$) , $Q_{8g}-Q_{8g}$, and $Q_{7\gamma}-Q_{8g}$ \cite{Benzke:2010js}. While not identified as resolved photon contributions, non-perturbative effects from $Q_{8g}-Q_{8g}$ \cite{Kapustin:1995fk} and $Q_1^q-Q_{7\gamma}$ \cite{Voloshin:1996gw,Ligeti:1997tc,Grant:1997ec,Buchalla:1997ky} were considered in the literature in the 1990's. Resolved photon contributions from $Q_{7\gamma}-Q_{8g}$ were only considered\footnote{See also \cite{Donoghue:1995na} for a model-dependent treatment.} in 2006 \cite{Lee:2006wn}. 

Resolved photon contributions have a more dramatic effect for the $\bar B\to X_s\gamma$ CP asymmetry due to the suppression of direct photon contributions in the SM \cite{Benzke:2010tq}. In the SM the CP asymmetry is dominated by the resolved photon contributions from $Q_1^q-Q_{7\gamma}$. Including both direct and resolved photon effects the SM prediction of \cite{Benzke:2010tq} is $-0.6\%<{\cal A}_{X_s\gamma}^{\rm SM}<2.8\%$. This can be compared to the 2019 update  of the 2018 PDG experimental value of $1.5\% \pm1.1\%$ \cite{Tanabashi:2018oca}. See also the HFLAV values in \cite{Amhis:2016xyh}.

In extensions of the SM in which there is a relative phase between the Wilson coefficients  $C_{7\gamma}$ and $C_1$ or $C_{7\gamma}$ and $C_{8g}$, resolved photon contributions lead to new CP-violating effects \cite{Benzke:2010tq}. In particular, effects from of $Q_{7\gamma}-Q_{8g}$ depends on the flavor of the spectator quark inside the $B$ meson. Such effects can be isolated by considering the difference between the CP asymmetries of charged and neutral $B$ mesons: $\Delta{\cal A}_{X_s\gamma} \equiv {\cal A}_{X_s^-\gamma} - {\cal A}_{X_s^0\gamma}$. In \cite{Benzke:2010tq} it was shown that $\Delta{\cal A}_{X_s\gamma}$ can reach the level of 10\% in magnitude for such models. This quantity was first measured by BaBar to be  $\Delta{\cal A}_{X_s\gamma}=(5.0\pm3.9\pm1.5)\%$ \cite{Lees:2014uoa} . Recently it was also measured by Belle to be  $\Delta{\cal A}_{X_s\gamma}=(3.69\pm2.65\pm0.76)\%$ \cite{Watanuki:2018xxg}. The PDG average of these results is $\Delta{\cal A}_{X_s\gamma}=(4.1\pm2.3)\%$ \cite{Tanabashi:2018oca}. 
The measurements are dominated by the statistical uncertainty. Therefore, the upcoming Belle II experiment is expected to reduce the uncertainty \cite{Watanuki:2018xxg}.

The analysis of \cite{Benzke:2010js} for the total rate gives the following contributions to the uncertainty from  resolved photon contributions. For  $Q_1^c-Q_{7\gamma}$ it was $[-1.7,+4.0]\,\%$. For $Q_{8g}-Q_{8g}$ it was $[-0.3,+1.9]\,\%$. For $Q_{7\gamma}-Q_{8g}$ two values were given. One based on vacuum insertion approximation (VIA) $[-2.8, -0.3]\,\%$ and the other based on experimental data available at the time $[-4.4, +5.6]\,\%$.  In particular the latter is related to $\Delta_{0-}$, the isospin asymmetry of inclusive neutral and charged $B$ decay to ${X_s\gamma}$. In 2010 only values from BaBar were available in \cite{Aubert:2005cua,Aubert:2007my}. 

Recently Belle has also published a value for $\Delta_{0-}$ \cite{Watanuki:2018xxg}. They obtain $\Delta_{0-}=[-0.48\pm1.49(\mbox{stat})\pm0.97 (\mbox{syst})\pm1.15(f_{+-}/f_{00})]\,\%$, where the last uncertainty is due to the uncertainty on the production ratio of $B^+B^-$ to $B^0\bar B^0$ in $\Upsilon(4S)$ decays.  The PDG average of \cite{Aubert:2005cua,Aubert:2007my,Watanuki:2018xxg} is $\Delta_{0-}=(-0.6\pm2.0)\%$. If we take the 95\% confidence level experimental range of this average, as was done in \cite{Benzke:2010js},  and translate it to the relative uncertainty via $-(1\pm0.3)\Delta_{0-}/3$  \cite{Misiak:2009nr,Benzke:2010js} we obtain that the $Q_{7\gamma}-Q_{8g}$ uncertainty is $[-1.4, +2]\,\%$.  This is more than  a factor of two reduction compared to the 2010 analysis \cite{Benzke:2010js}. 

Can we better constrain other resolved photon contributions?  For $Q_{8g}-Q_{8g}$ this seems difficult as little is known about the soft function it depends on.  For $Q_1^q-Q_{7\gamma}$, the contribution depends on a soft function whose moments are related to Heavy Quark Effective Theory (HQET) parameters. These HQET parameters appear also for inclusive semileptonic $B$ decays. The contributions to inclusive semileptonic $B$ decays of HQET parameters corresponding to dimension 7 and 8 HQET operators with tree level coefficients\footnote{When considering ${\cal O}(\alpha_s)$ coefficients, more operators appear \cite{Gunawardana:2017zix, Kobach:2017xkw}.} were classified in \cite{Mannel:2010wj}. In 2016 the authors of  \cite{Gambino:2016jkc} used moments of semileptonic $B$ decay spectra and information based on the lowest-lying state saturation approximation in \cite{Mannel:2010wj,Heinonen:2014dxa} to perform a global fit of these HQET parameters. Using the results of \cite{Gunawardana:2017zix} one can relate higher moments of the soft function to these HQET parameters. The goal of this paper is to use this new information to better constrain the $Q_1^q-Q_{7\gamma}$ resolved photon contribution to $\bar B\to X_s\gamma$  total rate and CP asymmetry.

The paper is structured as follows.  In section \ref{sec:moments} we derive general expressions for the moments of soft function in terms of HQET matrix elements. We relate such matrix elements of dimension 7 and 8 HQET operators to the HQET parameters of \cite{Mannel:2010wj}. In section \ref{sec:applications} we apply these relations to construct a systematic and improvable model of the soft function. We use the existing information about the HQET parameters to estimate the $Q_1^q-Q_{7\gamma}$ resolved photon contribution to $\bar B\to X_s\gamma$  total rate and CP asymmetry. We present our conclusions and outlook in section \ref{sec:conclusions}. In the appendix we derive a useful identity that is used in section \ref{sec:moments}.\\

\section{Moments of $\bm{g_{17}$}}\label{sec:moments}
The resolved photon contribution of $Q_1^q-Q_{7\gamma}$ to the $\bar B\to X_s\gamma$ spectrum depends on a non-perturbative soft function $g_{17}(\omega,\omega_1,\mu)$, see \cite{Benzke:2010js}. It is defined as
\begin{eqnarray}\label{g17def}
   g_{17}(\omega,\omega_1,\mu) 
   &=& \int\frac{dr}{2\pi}\,e^{-i\omega_1 r}\!
    \int\frac{dt}{2\pi}\,e^{-i\omega t} \\
   &&\times \frac{\langle\bar B| \big(\bar h S_n\big)(tn)\,
    \nbslash (1+\gamma_5) \big(S_n^\dagger S_{\bar n}\big)(0)\,
    i\gamma_\alpha^\perp\bar n_\beta\,
    \big(S_{\bar n}^\dagger\,g G_s^{\alpha\beta} S_{\bar n} 
    \big)(r\bar n)\,
    \big(S_{\bar n}^\dagger h\big)(0) |\bar B\rangle}{2M_B} \,,
    \nonumber
\end{eqnarray}
where $S_n$ and $S_{\bar n}$ are Wilson line defined in \cite{Benzke:2010js} and in the appendix. Intuitively we can think of $\omega$ as the momentum carried by the heavy quark and  $\omega_1$ as the momentum carried by the soft gluon. Moments of $g_{17}$ can be expressed in terms of HQET matrix elements and related to HQET parameters extracted from inclusive semileptonic $B$ decays. We consider three types of moments: moments in $\omega_1$ alone, moments in $\omega$ alone, and moments in both  $\omega_1$ and  $\omega$. We derive general expressions for each type of moments. We then relate matrix elements of HQET operators up to and including dimension 8 to known HQET parameters.     
\subsection{Moments in $\bm{\omega_1}$ alone}
We look at moments of the form 
\begin{eqnarray}
&&\langle\omega^0\,\omega_1^k\,g_{17}\rangle\equiv\int^{\bar \Lambda}_{-\infty}d\omega\int^{\infty}_{-\infty}d\omega_1\,\omega_1^k \,g_{17}(\omega,\omega_1,\mu)=\nonumber\\
&=&
(-1)^k\dfrac{1}{2M_B}\langle\bar B| \big(\bar h S_{\bar n}\big)(0)\,
    \nbslash (1+\gamma_5)\,
    i\gamma_\alpha^\perp\bar n_\beta\,
   (i\bar n\cdot \partial)^k \big(S_{\bar n}^\dagger\,g G_s^{\alpha\beta} S_{\bar n} 
    \big)(r\bar n)\,
    \big(S_{\bar n}^\dagger h\big)(0) |\bar B\rangle\big|_{r=0}.
\end{eqnarray}
Using the identity, proven in the appendix, $i\bar n\cdot \partial \left(S_{\bar n}^\dagger(x) O(x) S_{\bar n}(x) \right)=S_{\bar n}^\dagger(x)\big[i\bar n\cdot D,O(x)\big]S_{\bar n}(x)$, we can express the moments as matrix elements of the local operators 
\begin{eqnarray}\label{Generalw}
&&\langle\omega^0\,\omega_1^k\,g_{17}\rangle\equiv\int^{\bar \Lambda}_{-\infty}d\omega\int^{\infty}_{-\infty}d\omega_1\,\omega_1^k \,g_{17}(\omega,\omega_1,\mu)=\nonumber\\
&=&(-1)^k\dfrac{1}{2M_B}\langle\bar B| \bar h \,
    \nbslash (1+\gamma_5)\,
    i\gamma_\alpha^\perp\bar n_\beta\,\underbrace{\big[i\bar n\cdot D,\big[i\bar n \cdot D,\cdots[i\bar n \cdot D}_\text{$k$ times},g G_s^{\alpha\beta}\big]\cdots\big]\big] h |\bar B\rangle=\nonumber\\
&=&(-1)^k\dfrac{1}{2M_B}\langle\bar B| \bar h \,
    \nbslash (1+\gamma_5)\,
    \gamma_\alpha^\perp\,\underbrace{\big[i\bar n\cdot D,\big[i\bar n \cdot D,\cdots[i\bar n \cdot D}_\text{$k$ times},\big[iD^\alpha ,i\bar n \cdot D\big]\cdots\big]\big] h |\bar B\rangle.
\end{eqnarray}
 In the last line we have used the identity $\big[iD^\mu,iD^\nu\big]=igG^{\mu\nu}$. 
 
The Dirac structure of HQET matrix elements is simplified by using that $P_+h=h$, where $P_+\equiv(1+\vslash)/2$. As was shown in \cite{Mannel:1994kv}, between two $P_+$'s the Dirac basis reduces to four matrices: $P_+$ and $s^\lambda=P_+\gamma^\lambda\gamma^5P_+$. The matrices $s^\lambda$ are a generalization of the Pauli spin matrices that satisfy $v\cdot s=0$. This allows to simplify the Dirac structure.  

Consider $\nbslash\gamma_\alpha^\perp$ first. Since $\bar n$ and $\alpha$ are orthogonal, $\nbslash\gamma_\alpha^\perp=-i\sigma_{\mu\alpha_\perp}\bar n^\mu$. The matrix $(-i\sigma_{\mu\nu})$ is related to $s^\lambda$ via \cite{Mannel:1994kv} 
\begin{equation}
(-i\sigma_{\mu\nu})\to \frac{1+\vslash}{2}(-i\sigma^{\mu\nu})\frac{1+\vslash}{2}=iv^\rho\epsilon_{\rho\mu\nu\lambda}s^\lambda. 
\end{equation}
Note that this equation uses the convention $\epsilon_{0123}=-1$. Thus $\nbslash\gamma_\alpha^\perp\to iv^\rho\epsilon_{\rho\mu\alpha_\perp\lambda}s^\lambda\bar n^\mu$. The Dirac structure $\nbslash\gamma^5\gamma_\alpha^\perp$ can be simplified using the identity \cite{Mannel:1994kv} 
\begin{equation}
P_+\Gamma P_+=\dfrac12P_+\mbox{Tr}\left[P_+\Gamma\right]-\dfrac12 s^{\lambda}\mbox{Tr}\left[P_+ s_\lambda P_+\Gamma\right]\,,
\end{equation}
which gives $P_+\nbslash\gamma^5\gamma_\alpha^\perp P_+\to-s^{\alpha_\perp}$. We thus have 
\begin{eqnarray}
&&\langle\omega^0\omega_1^kg_{17}\rangle\equiv\int^{\bar \Lambda}_{-\infty}d\omega\int^{\infty}_{-\infty}d\omega_1\,\omega_1^k \,g_{17}(\omega,\omega_1,\mu)=\nonumber\\
&=&\left(iv^\rho\epsilon_{\rho\mu\alpha_\perp\lambda}\bar n^\mu-g_{\alpha_\perp\lambda}\right)(-1)^k\dfrac{1}{2M_B}\langle\bar B| \bar h \,
    \underbrace{\big[i\bar n\cdot D,\big[i\bar n \cdot D,\cdots[i\bar n \cdot D}_\text{$k$ times},\big[iD^\alpha ,i\bar n \cdot D\big]\cdots\big]\big] s^\lambda h |\bar B\rangle.\nonumber\\
\end{eqnarray}
The tensors in the last line can be related to the $g_\perp^{\mu\nu}$ and $\epsilon_\perp^{\mu\nu}$ defined in \cite{Bosch:2004cb}, although we will not need such relations. 

The nested commutator structure implies that odd moments vanish. The covariant derivative is a Hermitian operator. Odd number of commutators of Hermitian operators is a Hermitian operator whose forward matrix element is real. Since the spin-dependent matrix elements are imaginary, see \cite{Gunawardana:2017zix}, such matrix elements are zero. As was shown in \cite{Benzke:2010js}, the integral over $\omega$ of $g_{17}(\omega,\omega_1,\mu)$ is symmetric in $\omega_1$ which also implies that odd moments in $\omega_1$ must vanish.   

We  use the general decomposition of HQET matrix elements presented in \cite{Gunawardana:2017zix} to find the moments in $\omega_1$ up to the third moment. We find no contribution from the structure $\nbslash\gamma^5\gamma_\alpha^\perp$, as expected from  \cite{Benzke:2010js},  and that odd moments in $\omega_1$ vanish. We have 
\begin{eqnarray}\label{w1moments}
&&\langle\omega^0\,\omega_1^0\,g_{17}\rangle\equiv\int^{\bar \Lambda}_{-\infty}d\omega\int^{\infty}_{-\infty}d\omega_1\,g_{17}(\omega,\omega_1,\mu)=4\tilde{a}^{(5)}=2\lambda_2=2\mu_G^2/3\nonumber\\
&&\langle\omega^0\,\omega_1^1\,g_{17}\rangle\equiv\int^{\bar \Lambda}_{-\infty}d\omega\int^{\infty}_{-\infty}d\omega_1\,\omega_1 \,g_{17}(\omega,\omega_1,\mu)=0\nonumber\\
&&\langle\omega^0\,\omega_1^2\,g_{17}\rangle\equiv\int^{\bar \Lambda}_{-\infty}d\omega\int^{\infty}_{-\infty}d\omega_1\,\omega_1^2 \,g_{17}(\omega,\omega_1,\mu)=4 \left(-4 \tilde{a}^{(7)}_{12} + 2 \tilde{a}^{(7)}_{13} + 3 \tilde{a}^{(7)}_{14} - \tilde{a}^{(7)}_{23} + \tilde{b}^{(7)}\right)=\nonumber\\
&&=\dfrac2{15} \left(5 m_5 + 3 m_6 - 2 m_9\right)\nonumber\\
&&\langle\omega^0\,\omega_1^3\,g_{17}\rangle\equiv\int^{\bar \Lambda}_{-\infty}d\omega\int^{\infty}_{-\infty}d\omega_1\,\omega_1^3 \,g_{17}(\omega,\omega_1,\mu)=0.
\end{eqnarray}  
The zeroth moment is a known result. The result for the second moment is new.  

Here and in the following we are expressing the matrix elements in terms of the parameters $\tilde{a}^{(k)}_{ij}, \tilde{b}^{(k)}_{ij},\tilde{c}^{(k)}_{ij}$ of \cite{Gunawardana:2017zix}, $\lambda_2,\rho_2$ of \cite{Mannel:1994kv}, and $\mu_G^2,\rho^3_{LS}, m_i, r_i$ of \cite{Mannel:2010wj}. See \cite{Gunawardana:2017zix} for definitions and relations between these parameters.  Since $\tilde{a}^{(k)}_{ij}, \tilde{b}^{(k)}_{ij},\tilde{c}^{(k)}_{ij}$ and $\lambda_2,\rho_2$ are defined in the heavy quark limit while $\mu_G^2,\rho^3_{LS}, m_i, r_i$ are not, there are $1/m_b$ differences between, e.g., $\lambda_2$ and $\mu_G^2/3$. We discuss these differences  in section \ref{subsec:numerical}.

\subsection{Moments in $\bm{\omega}$ alone}
We look at moments of the form 
\begin{eqnarray}\label{Generalw1}
&&\langle\omega^k\,\omega_1^0\,g_{17}\rangle\equiv\int^{\bar \Lambda}_{-\infty}d\omega\,\omega^k \int^{\infty}_{-\infty}d\omega_1\,g_{17}(\omega,\omega_1,\mu)=\nonumber\\
&=&\int^{\bar \Lambda}_{-\infty}d\omega\,\omega^k  \int\frac{dt}{2\pi}\,e^{-i\omega t} 
\dfrac{1}{2M_B}\langle\bar B| \big(\bar h S_{n}\big)(tn)\,
    \nbslash (1+\gamma_5)\,S_n^\dagger (0)\,
    i\gamma_\alpha^\perp\bar n_\beta\,
   g G_s^{\alpha\beta}(0)
    h(0) |\bar B\rangle=\nonumber\\
&=&\int^{\bar \Lambda}_{-\infty}d\omega\,\omega^k  \int\frac{dt}{2\pi}\,e^{i\omega t} 
\dfrac{1}{2M_B}\langle\bar B| \big(\bar h S_{n}\big)(0)\,
    \nbslash (1+\gamma_5)\,S_n^\dagger (tn)\,
    i\gamma_\alpha^\perp\bar n_\beta\,
   g G_s^{\alpha\beta}(tn)
    h(tn) |\bar B\rangle=\nonumber\\
&=&\int\,dt\,\delta(t)
\dfrac{1}{2M_B}\langle\bar B|\bar h(0) S_{n}(0)\,(in\cdot\partial)^kS_n^\dagger (tn)\,
    \nbslash (1+\gamma_5)\,
    i\gamma_\alpha^\perp\bar n_\beta\,
   g G_s^{\alpha\beta}(tn)
    h(tn) |\bar B\rangle.
        \end{eqnarray}
where we have used the translation invariance of forward matrix elements of non-local operators.  The identity $S_{n}^\dagger(x)\,in\cdot D\,S_n(x)=in\cdot \partial$ which follows from $in\cdot D\,S_{n}(x)=0$ implies that $S_{n}^\dagger(tn)\,in\cdot D\,=in\cdot \partial \,S^\dagger_n(tn)$. This allows us to express the moments as 
\begin{eqnarray}
&&\langle\omega^k\,\omega_1^0\,g_{17}\rangle\equiv\int^{\bar \Lambda}_{-\infty}d\omega\,\omega^k \int^{\infty}_{-\infty}d\omega_1\,g_{17}(\omega,\omega_1,\mu)=\nonumber\\
&=&\left(iv^\rho\epsilon_{\rho\mu\alpha_\perp\lambda}\bar n^\mu-g_{\alpha_\perp\lambda}\right)\dfrac{1}{2M_B}\langle\bar B| \bar h \,
   \left(in \cdot D\right)^k\big[iD^\alpha ,i\bar n \cdot D\big]s^\lambda h |\bar B\rangle.
\end{eqnarray}
Notice that the location of  $\left(in \cdot D\right)^k$ is determined by the Wilson lines in the $n$ direction. 

We use the general decomposition of HQET matrix elements presented in \cite{Gunawardana:2017zix} to find the moments in $\omega$ up to the third moment. As before there is no contribution from the structure $\nbslash\gamma^5\gamma_\alpha^\perp$.  We have 
\begin{eqnarray}
&&\langle\omega^0\,\omega_1^0\,g_{17}\rangle\equiv\int^{\bar \Lambda}_{-\infty}d\omega\int^{\infty}_{-\infty}d\omega_1\,g_{17}(\omega,\omega_1,\mu)=4\tilde{a}^{(5)}=2\lambda_2=2\mu_G^2/3\nonumber\\
&&\langle\omega^1\,\omega_1^0\,g_{17}\rangle\equiv\int^{\bar \Lambda}_{-\infty}d\omega\,\omega\int^{\infty}_{-\infty}d\omega_1 \,g_{17}(\omega,\omega_1,\mu)=-2\tilde{a}^{(6)}=-\rho_2=-\rho^3_{LS}/3\nonumber\\
&&\langle\omega^2\,\omega_1^0\,g_{17}\rangle\equiv\int^{\bar \Lambda}_{-\infty}d\omega\,\omega^2\int^{\infty}_{-\infty}d\omega_1 \,g_{17}(\omega,\omega_1,\mu)=-2 \left(2 \tilde{a}^{(7)}_{12} - \tilde{a}^{(7)}_{14} +  \tilde{a}^{(7)}_{23} + \tilde{b}^{(7)}\right)=\nonumber\\
&&=-\dfrac1{60} \left(20 m_5 +2 m_7 + m_8\right)\nonumber\\
&&\langle\omega^3\,\omega_1^0\,g_{17}\rangle\equiv\int^{\bar \Lambda}_{-\infty}d\omega\,\omega^3\int^{\infty}_{-\infty}d\omega_1 \,g_{17}(\omega,\omega_1,\mu)=\nonumber\\
&&=-2 \left(2 \tilde{a}^{(8)}_{12} - \tilde{a}^{(8)}_{15} +  \tilde{a}^{(8)}_{24} +2 \tilde{b}^{(8)}_{13} + \tilde{b}^{(8)}_{14} - \tilde{b}^{(8)}_{15} - 2\tilde{b}^{(8)}_{35} - \tilde{b}^{(8)}_{45} + \tilde{c}^{(8)}\right)=\nonumber\\
&&=-\dfrac1{15} \left(5r_8 - r_9 +2 r_{10}+r_{11}-2r_{12}-r_{13}+2r_{15}-r_{16}+r_{17}\right)
\end{eqnarray}  
The first moment was derived in \cite{Benzke:2010js}. The results for the second and third moments are new.

\subsection{Moments in both $\bm{\omega_1}$ and $\bm{\omega}$}
Combining the derivations in the previous subsections, the expression for the mixed moments in $\omega_1$ and $\omega$ is 
\begin{eqnarray}\label{Generalww1}
&&\langle\omega^l\,\omega_1^k\,g_{17}\rangle\equiv\int^{\bar \Lambda}_{-\infty}d\omega\,\omega^l \int^{\infty}_{-\infty}d\omega_1\,\omega^k\,g_{17}(\omega,\omega_1,\mu)=\left(iv^\rho\epsilon_{\rho\mu\alpha_\perp\lambda}\bar n^\mu-g_{\alpha_\perp\lambda}\right)(-1)^k\times\nonumber\\
&\times&\dfrac{1}{2M_B}\langle\bar B| \bar h \,(i n\cdot D)^l
    \underbrace{\big[i\bar n\cdot D,\big[i\bar n \cdot D,\cdots[i\bar n \cdot D}_\text{$k$ times},\big[iD^\alpha ,i\bar n \cdot D\big]\cdots\big]\big] s^\lambda h |\bar B\rangle.
\end{eqnarray}  
We use the general decomposition of HQET matrix elements presented in \cite{Gunawardana:2017zix} to find the mixed moments in $\omega$ up to operators of dimension 8. These are 
\begin{eqnarray}
&&\langle\omega^1\,\omega_1^1\,g_{17}\rangle\equiv\int^{\bar \Lambda}_{-\infty}d\omega\,\omega\int^{\infty}_{-\infty}d\omega_1\,\omega_1 \,g_{17}(\omega,\omega_1,\mu)=2 \left(-4 \tilde{a}^{(7)}_{12}+ 2\tilde{a}^{(7)}_{13} +3 \tilde{a}^{(7)}_{14} -  \tilde{a}^{(7)}_{23} + \tilde{b}^{(7)}\right)=\nonumber\\
&&=\dfrac1{15} \left(5 m_5 +3 m_6 -2m_9\right)\nonumber\\
&&\langle\omega^2\,\omega_1^1\,g_{17}\rangle\equiv\int^{\bar \Lambda}_{-\infty}d\omega\,\omega^2\int^{\infty}_{-\infty}d\omega_1\,\omega_1 \,g_{17}(\omega,\omega_1,\mu)=\nonumber\\
&&=2 \left(3 \tilde{a}^{(8)}_{12} - \tilde{a}^{(8)}_{14}  -2\tilde{a}^{(8)}_{15} +  \tilde{a}^{(8)}_{24} -3 \tilde{b}^{(8)}_{13} + \tilde{b}^{(8)}_{14} +4\tilde{b}^{(8)}_{15} +3\tilde{b}^{(8)}_{35} - \tilde{b}^{(8)}_{45} + \tilde{c}^{(8)}\right)=\nonumber\\
&&=\dfrac1{15} \left(5r_8 - r_9 -3 r_{10}+r_{11}+3r_{12}+4r_{13}+3r_{15}-2r_{16}+r_{17}-r_{18}\right)\nonumber\\
&&\langle\omega^1\,\omega_1^2\,g_{17}\rangle\equiv\int^{\bar \Lambda}_{-\infty}d\omega\,\omega^1\int^{\infty}_{-\infty}d\omega_1\,\omega_1^2 \,g_{17}(\omega,\omega_1,\mu)=\nonumber\\
&&=2 \left(3 \tilde{a}^{(8)}_{12} - \tilde{a}^{(8)}_{14} - 2\tilde{a}^{(8)}_{15} +  \tilde{a}^{(8)}_{24} +3 \tilde{b}^{(8)}_{13} + \tilde{b}^{(8)}_{14} - 2\tilde{b}^{(8)}_{15} + 2\tilde{b}^{(8)}_{34} -\tilde{b}^{(8)}_{35} + \tilde{b}^{(8)}_{45} - \tilde{c}^{(8)}\right)=\nonumber\\
&&=\dfrac1{15} \left(-5r_8 + r_9 +3 r_{10}+r_{11}-r_{12}-2r_{13}+2r_{14}+3r_{15}-2r_{16}+r_{17}-r_{18}\right)
\end{eqnarray}  
As before, there is no contribution from the structure $\nbslash\gamma^5\gamma_\alpha^\perp$. All the results for these moments are new.    

\section{Applications}\label{sec:applications}
\subsection{Current numerical values of moments}\label{subsec:numerical}
The HQET parameters arising from matrix elements of HQET operators up to dimension 8 were extracted from experimental data in 2016 \cite{Gambino:2016jkc}. The authors of  \cite{Gambino:2016jkc} used moments of semileptonic $B$ decay spectra and information based on the lowest-lying state saturation approximation  in \cite{Mannel:2010wj,Heinonen:2014dxa} to perform a global fit of these HQET parameters. Based on the values and standard deviations given in Table 2 of \cite{Gambino:2016jkc} the non-zero moments of $g_{17}$ are
\begin{eqnarray}\label{moments_numerical}
\langle\omega^0\,\omega_1^0\,g_{17}\rangle&=&0.237\pm0.040 \mbox{ GeV}^2\nonumber\\
\langle\omega^0\,\omega_1^2\,g_{17}\rangle&=&0.15\pm0.12 \mbox{ GeV}^4\nonumber\\
\langle\omega^1\,\omega_1^0\,g_{17}\rangle&=&0.056\pm0.032 \mbox{ GeV}^3\nonumber\\
\langle\omega^2\,\omega_1^0\,g_{17}\rangle&=&0.015\pm0.021 \mbox{ GeV}^4\nonumber\\
\langle\omega^3\,\omega_1^0\,g_{17}\rangle&=&0.008\pm0.011 \mbox{ GeV}^5\nonumber\\
\langle\omega^1\,\omega_1^1\,g_{17}\rangle&=&0.073\pm0.059 \mbox{ GeV}^4\nonumber\\
\langle\omega^2\,\omega_1^1\,g_{17}\rangle&=&-0.034\pm0.016 \mbox{ GeV}^5\nonumber\\
\langle\omega^1\,\omega_1^2\,g_{17}\rangle&=&0.027\pm0.014 \mbox{ GeV}^5,
\end{eqnarray}  
where we have added the error bars of individual HQET parameters in quadrature. We do not include correlations as none were given in Table 2 of \cite{Gambino:2016jkc}.   

\begin{figure}
\begin{center}
\includegraphics[scale=0.6]{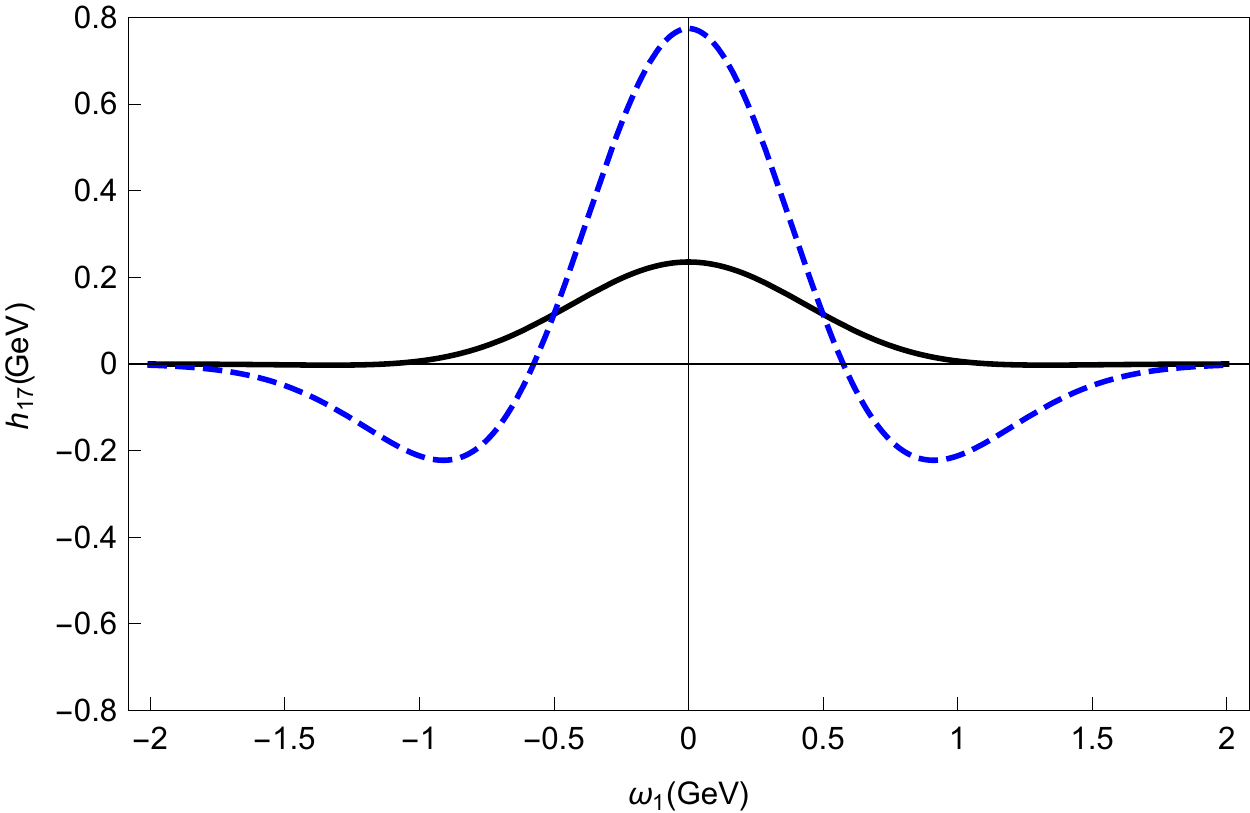}
\hspace{1cm}
\includegraphics[scale=0.6]{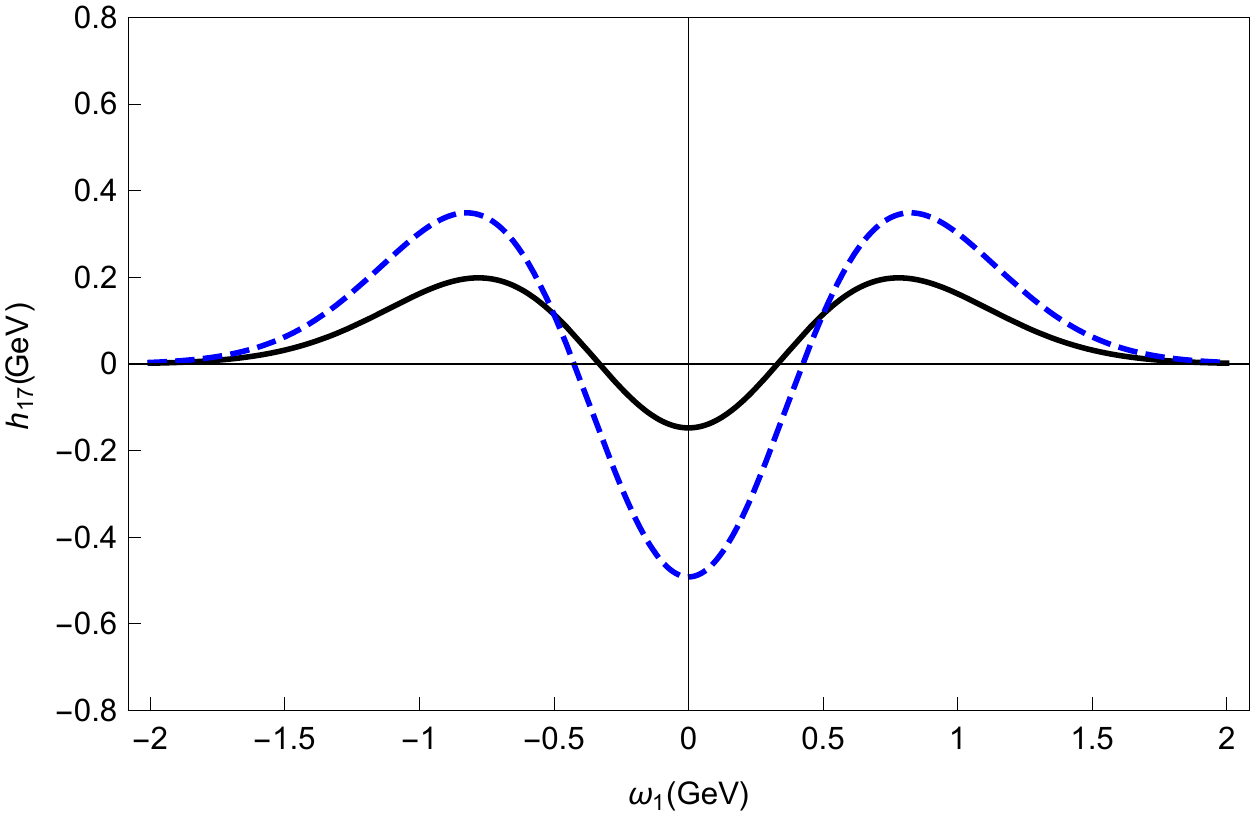}
\caption{\label{fig1} A comparison of the extremal models for $h_{17}$ as a sum of two lowest even Hermite polynomials times a Gaussian of width $0.5$ GeV used in \cite{Benzke:2010js} (dashed blue) to the same models allowed by current (2019) data (solid black). Left hand side:  The model with 2010 smallest possible second moment of $-0.31\mbox{ GeV}^4$ compared to 2019 smallest possible second moment of $0.03\mbox{ GeV}^4$. Right hand side:  The model with 2010 largest possible second moment of $0.49\mbox{ GeV}^4$ compared to 2019 largest possible second moment of $0.27\mbox{ GeV}^4$. }
\end{center}
\end{figure}

While the relative errors are large, the moments do give useful information. For example, the two extremal models used  in \cite{Benzke:2010js} for $h_{17}$, defined in (\ref{h17def}), have $\langle\omega^0\,\omega_1^2\,g_{17}\rangle=-0.31\mbox{ GeV}^4$ and $\langle\omega^0\,\omega_1^2\,g_{17}\rangle=0.49\mbox{ GeV}^4$. Using the value above of $0.15\pm0.12 \mbox{ GeV}^4$ this corresponds to  roughly a three standard deviations range, as opposed to the one standard deviation range in (\ref{moments_numerical}). Similarly, in figure \ref{fig1} we compare the models of \cite{Benzke:2010js} for $h_{17}$ that used a sum of two Hermite polynomials to the sum of two Hermite polynomial model for $h_{17}$ (defined below) with the current extremal values of $\langle\omega^0\,\omega_1^2\,g_{17}\rangle$. 

As was alluded to in the previous section, the parameters defined in \cite{Mannel:2010wj} and listed in \cite{Gambino:2016jkc} use the full QCD $b$ fields, while the matrix elements we need are defined in the heavy quark limit. This implies that there are $1/m_b$ differences between, e.g., $\lambda_2$ and $\mu_G^2/3$, see \cite{Gremm:1996df,Neubert:2005nt}. Since the error bars in Table 2 of \cite{Gambino:2016jkc} are rather large, one might question if this issue is numerically important. To test this, we compare the value of $\mu_G^2/3$ from \cite{Gambino:2016jkc} to extractions of $\lambda_2$ (which is defined in the heavy quark limit) from $B$ and $D$ meson spectroscopy.

We define $\Delta m_H=m_H^*-m_H$, where $m_H$ ($m_H^*$) is a pseudo-scalar (vector) heavy meson containing a heavy quark of mass $m_Q$. The expression for $\Delta m_H$ up to order $1/m_Q^2$ can be found\footnote{For consistency with the rest of our paper, we do not include the scale dependance of $\lambda_2$ which is an ${\cal O}\left(\alpha_s\right)$ effect.}  in \cite{Gremm:1996df}. To extract $\lambda_2$ we use isospin-averaged meson mass data from the 2019 update of the 2018 PDG review \cite{Tanabashi:2018oca}. At order $1/m_Q$, $\lambda_2=\Delta m_Hm_H/2$. Thus $\lambda_2=0.119\pm 0.001\mbox{ GeV}^2$ from $B$-meson data, and  $\lambda_2=0.13193\pm0.00002\mbox{ GeV}^2$ from $D$-meson data. The errors are only from the meson masses.  At order $1/m_Q^2$, $\lambda_2=(\Delta m_Bm_B^2-\Delta m_Dm_D^2)/(2m_B-2m_D)$ \cite{Gremm:1996df}. Thus $\lambda_2=0.112\pm 0.001\mbox{ GeV}^2$.  Comparing to $\mu_G^2/3=0.118\pm 0.020\mbox{ GeV}^2$  \cite{Gambino:2016jkc}, we see that it is equal to all of these values of $\lambda_2$  within errors.  Thus currently we cannot distinguish the two. To be conservative, we will use  $\mu_G^2/3$ from \cite{Gambino:2016jkc}. We assume that a similar situation applies to other HQET parameters.

One would expect that in the future data from Belle II or Lattice QCD will allow to further constrain the HQET parameters and hence the moments of $g_{17}$.

\subsection{Resolved photon contributions for $\bm{Q^q_{1}-Q_{7\gamma}}$}\label{subsec:resolved}
The information about the moments presented above can be used to better constrain the resolved photon contribution of $Q^q_{1}-Q_{7\gamma}$. The observables we consider are the CP averaged rate and the CP asymmetry, both integrated over the photon energy $E_0\leq E_\gamma\leq M_B/2$ where $M_B$ is the $B$ meson mass. As was discussed in \cite{Benzke:2010js}, provided that $\Delta\equiv m_b-2E_0$ is much larger than $\Lambda_{\mbox{\scriptsize QCD}}$, the contribution of $Q^q_{1}-Q_{7\gamma}$ is expressed in terms of the soft function 
\begin{equation}\label{h17def}
h_{17}(\omega_1,\mu)=  \int\frac{dr}{2\pi}\,e^{-i\omega_1 r} \frac{\langle\bar B| \big(\bar h S_{\bar n}\big)(0)\,
    \nbslash \,
    i\gamma_\alpha^\perp\bar n_\beta\,
    \big(S_{\bar n}^\dagger\,g G_s^{\alpha\beta} S_{\bar n} 
    \big)(r\bar n)\,
    \big(S_{\bar n}^\dagger h\big)(0) |\bar B\rangle}{2M_B} \,.
\end{equation}
obtained from $g_{17}(\omega,\omega_1,\mu)$ by integrating over $\omega$ and omitting $\gamma^5$ \cite{Benzke:2010js}.

For the CP averaged rate the quantity we are interested in is ${\cal F}_E(\Delta)$ corresponding to the relative theoretical uncertainty  from the  resolved photon contributions. As shown in \cite{Benzke:2010js} its $Q^u_{1}-Q_{7\gamma}$ part vanishes. Its $Q^c_{1}-Q_{7\gamma}$ part is 
\begin{equation}\label{F17def}
{\cal F}^{17}_E = \frac{C_1(\mu)}{C_{7\gamma}(\mu)}\,\frac{\Lambda_{17}(m_c^2/m_b,\mu)}{m_b},
\end{equation}
where 
\begin{equation}\label{Lambda17def}
   \Lambda_{17}\Big(\frac{m_c^2}{m_b},\mu\Big)
   = e_c\,\mbox{Re} \int_{-\infty}^\infty \frac{d\omega_1}{\omega_1} 
    \left[ 1 - F\!\left( \frac{m_c^2-i\varepsilon}{m_b\,\omega_1} \right)
    + \frac{m_b\,\omega_1}{12m_c^2} \right] h_{17}(\omega_1,\mu)\,,
\end{equation}
and $F(x) = 4x\arctan^2\left(1/{\sqrt{4x-1}} \right)$. Assuming $\Delta\gg \Lambda_{\mbox{\scriptsize QCD}}$ allows to replace $g_{17}(\omega,\omega_1,\mu)$ by $h_{17}(\omega_1,\mu)$ and ignore the $\Delta$ dependence in ${\cal F}^{17}_E(\Delta)$.  In \cite{Benzke:2010js} $ \Lambda_{17}$ was estimated to be in the range $-60 \mbox{ MeV} <  \Lambda_{17} < 25 \mbox{ MeV}$. We would like to use the information about the moments to revisit this estimate.  

For the CP asymmetry the quantity we are interested in is the $Q^q_{1}-Q_{7\gamma}$ part of the resolved photon contribution to the CP asymmetry ${\cal A}_{X_s\gamma}^{\rm res,17}$, defined as \cite{Benzke:2010tq} 
\begin{equation}
   {\cal A}_{X_s\gamma}^{\rm res,17}
   = \frac{\pi}{m_b}\,\bigg\{
    \mbox{Im}\bigg[(1+\epsilon_s)\,\frac{C_1}{C_{7\gamma}}\bigg]\,\tilde\Lambda_{17}^c 
    - \mbox{Im}\bigg[\epsilon_s\,\frac{C_1}{C_{7\gamma}}\bigg]\,\tilde\Lambda_{17}^u\bigg\}, 
    \end{equation}
where 
\begin{equation}\label{TildeLambda17def}
   \tilde\Lambda_{17}^u = \frac23\,h_{17}(0) \,, \quad
   \tilde\Lambda_{17}^c 
   = \frac23 \int_{4m_c^2/m_b}^\infty\!\frac{d\omega}{\omega}\,
    f\bigg( \frac{m_c^2}{m_b\,\omega} \bigg)\,h_{17}(\omega) \,,
\end{equation}
with 
\begin{equation}
   f(x) = 2x\ln\frac{1+\sqrt{1-4x}}{1-\sqrt{1-4x}} \,.
\end{equation}
In \cite{Benzke:2010tq} $ \tilde\Lambda_{17}^u$ and  $\tilde\Lambda_{17}^c$ were estimated to be in the range $- 330\,\mbox{MeV} < \tilde\Lambda_{17}^u < + 525\,\mbox{MeV}$ and $- 9\,\mbox{MeV} < \tilde\Lambda_{17}^c < + 11\,\mbox{MeV}$. We would like to use the information about the moments to revisit these estimates.  

To do that, we will consider various approaches to estimate the hadronic uncertainty arising from the soft function $h_{17}$ and its moments. We also take into account the uncertainty from the values of the charm and bottom quark masses. They appear in the functions $F(x)$ and $f(x)$ above. 

As discussed in \cite{Benzke:2010js}, the charm-quark mass enters as a running mass in charm-penguin diagrams with a soft gluon emission, which are characterized by a hard-collinear virtuality.  As in \cite{Benzke:2010js} we use $m_c = m_c(\mu)$ defined in the $\overline{\mbox{MS}}$ scheme with $\mu=1.5$ GeV for the CP averaged rate. As in \cite{Benzke:2010tq} we use $m_c = m_c(\mu)$ defined in the $\overline{\mbox{MS}}$ scheme with $\mu=2.0$ GeV for the CP asymmetry.  We will comment on the choice of scales in the conclusions.

The 2019 update of the 2018 PDG listing has $m_c(\,m_c)=1.27\pm0.02$ GeV \cite{Tanabashi:2018oca}. This is an average of masses in other schemes converted to the $\overline{\mbox{MS}}$ scheme using two-loop QCD perturbation theory with $\alpha_s(\mu=m_c) = 0.38 \pm0.03$ \cite{Tanabashi:2018oca}. We use the same two-loop QCD perturbation theory \cite{Buras:1998raa} to find $m_c(1.5 \mbox{ GeV})=1.20\pm0.03$ GeV and $m_c(2.0 \mbox{ GeV})=1.10\pm0.03$ GeV. This should be compared, for example,  to the value used in \cite{Benzke:2010js} of $1.131$ GeV derived based on a smaller value of $m_c(m_c)$ from \cite{Hoang:2005zw} that was also used in \cite{Misiak:2006zs,Misiak:2006ab}. The change in the value of the charm quark mass tends to slightly change the size of  $\Lambda_{17}$ and $\tilde\Lambda_{17}^c$. 

Following \cite{Benzke:2010js} we will use the value of the bottom quark in the shape function scheme \cite{Bosch:2004th}. The latest HFLAV \cite{Amhis:2016xyh} value is $m_b=4.58\pm0.03$ GeV. This should be compared to the value of $4.65$ GeV used in \cite{Benzke:2010js}.

\subsection{$\bm{\Lambda_{17}}$ estimates based on expanded penguin function}\label{subsec:expanded}
The soft function $h_{17}$ appears in $\Lambda_{17}$  convoluted with a penguin function $F$ that depends on the ratio $m_c^2$ to the anti-hard-collinear scale $m_b\,\omega_1$. For $x>1/4$, $1-F(x)$ has the expansion
\begin{equation}\label{PenguinExpanded} 
1-F(x)=-\dfrac1{12x}-\dfrac1{90x^2}-\dfrac1{560x^3}+{\cal O}\left(\dfrac1{x^4}\right)
\end{equation} 

\emph{Assuming} that $h_{17}$ has support for values of $\omega_1\ll 4m^2_c/m_b\approx 1.2-1.3$ GeV we can expand the penguin function and express $\Lambda_{17}$ in terms of the moments of $h_{17}$. From the definition of $h_{17}$ it is clear that $\langle\omega^0\,\omega_1^k\,g_{17}\rangle=\langle\omega_1^k\,h_{17}\rangle$. Thus
\begin{equation}\label{Lambda17expanded}
\Lambda_{17}^{\mbox{\scriptsize expanded}}=  -\dfrac{e_c m_b^3}{560\,m_c^6} \langle\omega^0\,\omega_1^2\,g_{17}\rangle+\cdots=-6\pm5\mbox{ MeV}+\cdots,
\end{equation}
where $\cdots$ denotes the contribution of higher moments in $\omega_1$. Odd moments in $\omega_1$ vanish. The contribution of the zeroth moment in $\omega_1$ is subtracted in (\ref {Lambda17def}) since it is traditionally not included in the resolved photon contributions. Its size is  $-e_cm_b2\lambda_2/(12m_c^2)=-42\pm7 \mbox{ MeV}$.  The uncertainty  in (\ref{Lambda17expanded}) arises from $\langle\omega^0\,\omega_1^2\,g_{17}\rangle$, $m_b$, and $m_c$ added in quadrature. 

The size of the contribution of higher dimensional operators was a concern for the authors of \cite{Voloshin:1996gw,Ligeti:1997tc,Grant:1997ec,Buchalla:1997ky}. They have noticed the numerical suppression arising from the expansion of the penguin function, see (\ref{PenguinExpanded}), but the lack of knowledge of the matrix elements prevented them from making conclusive statements. The new numerical information about the higher dimensional matrix elements allows us to address this issue for the first time.  

The expansion of the penguin function generates a numerical suppression factor of $\sim50$ between the first and third term\footnote{Recall that the second term combines with $\langle\omega^0\,\omega_1\,g_{17}\rangle$ which is zero.} in (\ref{PenguinExpanded}). Despite that, when combined with the second moment, the central value of the total contribution in (\ref{Lambda17expanded}), $-6\mbox{ MeV}$, is only suppressed by a factor of seven compared to the central value of the contribution proportional to the zeroth moment, $-42 \mbox{ MeV}$.  The smaller suppression is consistent with the power counting of $m_c^2\sim m_b\Lambda_{\mbox{\scriptsize QCD}}$ which disfavors the expansion of the penguin function. 

As in \cite{Benzke:2010js}, it is instructive to look at $\Lambda_{\mbox{\scriptsize QCD}}/m_b$ corrections to $\Lambda_{17}^{\mbox{\scriptsize expanded}}$. In \cite{Benzke:2010js} only the moment $\langle\omega\,\omega_1^0\,g_{17}\rangle$ was considered. We are at the position now to consider other moments too. The starting point is the expression for $\Lambda_{17}$ that includes the photon energy dependance beyond leading power in $\Lambda_{\mbox{\scriptsize QCD}}/m_b$ \cite{Benzke:2010js} 
\begin{eqnarray}\label{L17power}
   \Lambda_{17}\Big(\frac{m_c^2}{m_b},\mu\Big)
   &=& e_c\,\mbox{Re} \int_{-\infty}^{\bar\Lambda}\!d\omega
    \int_{-\infty}^\infty \frac{d\omega_1}{\omega_1} \nonumber\\
   &\quad\times& \left\{ \left( \frac{m_b+\omega}{m_b} \right)^3 
    \left[ 1 - F\!\left( \frac{m_c^2-i\varepsilon}{(m_b+\omega)\,\omega_1} \right) \right]
    + \frac{m_b\,\omega_1}{12m_c^2} \right\} g_{17}(\omega,\omega_1,\mu) \,.
\end{eqnarray}
We expand $F(x)$ as above and in $\omega/m_b$ and consider $1/m_b^n$ corrections to $\Lambda_{17}^{\mbox{\scriptsize expanded}}$ denoted by $\delta \Lambda_{17}^{(n)}$. By definition, $\Lambda_{17}^{\mbox{\scriptsize expanded}}=\delta \Lambda_{17}^{(0)}$. For $\delta\Lambda_{17}^{(1)}$ we have 
\begin{eqnarray}
\delta \Lambda_{17}^{(1)}&=& -\dfrac{e_c}{3\,m_c^2} \langle\omega^1\,\omega_1^0\,g_{17}\rangle-\dfrac{e_cm_b}{18\,m_c^4} \langle\omega^1\,\omega_1^1\,g_{17}\rangle-\dfrac{3e_cm_b^2}{280\,m_c^6} \langle\omega^1\,\omega_1^2\,g_{17}\rangle+\cdots\nonumber\\
&=&\left(-9\pm5\mbox{ MeV}\right)+\left(-6\pm5\mbox{ MeV}\right)+\left(-1\pm1\mbox{ MeV}\right)+\cdots=-16\pm7\mbox{ MeV}+\cdots\,.\nonumber\\ 
\end{eqnarray}
We notice again a slow convergence in the series generated from the expansion of $F(x)$. Only in the third term can we see a suppression compared to the pervious terms. Although nominally a $\Lambda_{\mbox{\scriptsize QCD}}/m_b$ correction, $\delta \Lambda_{17}^{(1)}$ is comparable in size to $\Lambda_{17}^{\mbox{\scriptsize expanded}}$. Even if we add the contribution of $\langle\omega^0\,\omega_1^0\,g_{17}\rangle$ to $\Lambda_{17}^{\mbox{\scriptsize expanded}}$, $\delta \Lambda_{17}^{(1)}$ is only suppressed by a factor of  three. 
The $\Lambda^2_{\mbox{\scriptsize QCD}}/m^2_b$ correction for  $\Lambda_{17}^{\mbox{\scriptsize expanded}}$ is 
\begin{eqnarray}
\delta \Lambda_{17}^{(2)}&=& -\dfrac{e_c}{2\,m_b\,m_c^2} \langle\omega^2\,\omega_1^0\,g_{17}\rangle-\dfrac{e_c}{9\,m_c^4} \langle\omega^2\,\omega_1^1\,g_{17}\rangle+\cdots\nonumber\\
&=&\left(-0.8\pm1.1\mbox{ MeV}\right)+\left(1.2\pm0.6\mbox{ GeV}\right)+\cdots=0.4\pm1.3\mbox{ MeV}+\cdots\,.
\end{eqnarray}
Again we observe a slow convergence in the series generated from the expansion of $F(x)$. The overall magnitude in this case is consistent with a simple  $\Lambda_{\mbox{\scriptsize QCD}}/m_b\sim0.1$ expectation. Finally, the $\Lambda^3_{\mbox{\scriptsize QCD}}/m^3_b$ correction for  $\Lambda_{17}^{\mbox{\scriptsize expanded}}$ is  
\begin{eqnarray}
\delta \Lambda_{17}^{(3)}&=& -\dfrac{e_c}{3\,m_b^2\,m_c^2} \langle\omega^3\,\omega_1^0\,g_{17}\rangle=-0.06\pm0.08\mbox{ MeV}+\cdots\,.
\end{eqnarray}
As for $\delta \Lambda_{17}^{(2)}$, we observe the expected order of magnitude reduction in going to the next term in $\Lambda_{\mbox{\scriptsize QCD}}/m_b$.  

We see that numerically the $\Lambda_{\mbox{\scriptsize QCD}}/m_b$ expansion for $\delta \Lambda_{17}$ works well with the exception of the first term. One can speculate that the vanishing of $\langle\omega^0\,\omega_1^1\,g_{17}\rangle$ makes the zeroth term in the expansion $\Lambda_{17}$ of (\ref{L17power}) smaller than it ``should" be. Since in general for $l>0$ the moments $\langle\omega^l\,\omega_1^k\,g_{17}\rangle$ do not vanish, there is no such suppression beyond the zeroth term. Adding the terms above linearly and their uncertainties in quadrature gives $\Lambda_{17}^{\mbox{\scriptsize expanded}}+\delta \Lambda_{17}^{(1)}+\delta \Lambda_{17}^{(2)}+\delta \Lambda_{17}^{(3)}=-22\pm9\mbox{ MeV}$.

As was discussed in \cite{Benzke:2010js}, the assumptions about the support of $h_{17}$ and the resulting expansion of the penguin function are too restrictive.  We turn to estimates that do not relay on this expansion.

\subsection{Modeling of $\bm{h_{17}}$}
As was shown in \cite{Benzke:2010js}, $h_{17}(\omega_1,\mu)$ is an even function. It also has a dimension of mass and in the heavy quark limit $-\infty\leq\omega_1\leq \infty$. In modeling $h_{17}$ it is beneficial to have a systematic expansion of $h_{17}$, e.g. in terms of a complete orthonormal set of basis functions. For the leading power shape function such an expansion was suggested in \cite{Ligeti:2008ac}. We will use an expansion in terms of Hermite polynomials multiplied by a Gaussian of width $\sigma$:
\begin{equation}\label{h17General} 
h_{17}(\omega_1,\mu)=\sum_n a_{2n} H_{2n}\left(\frac{\omega_1}{\sqrt{2}\sigma}\right)e^{-\frac{\omega_1^2}{2\sigma^2}}. 
\end{equation}
Since $h_{17}$ is even, only even polynomials are needed. In the following we refer to these models by the numbers of Hermite polynomials they contain. Since the Hermite polynomials are orthogonal, the  $2k$-th moment of $h_{17}$ only depends on the coefficients $a_{2n}$ with $n\leq k$, for a given value of $\sigma$. In other words, the lack of knowledge of higher moments does not affect models that only use lower moments. For example, the zeroth moment of $h_{17}$ only depends on $a_0$ and the second moment of $h_{17}$ only depends on $a_0$ and $a_2$. Conversely, we can use the first $2k$-th moments to determine $a_{2n}$ with $n\leq k$. Using $\langle\omega^0\,\omega_1^k\,g_{17}\rangle=\langle\omega_1^k\,h_{17}\rangle$ we have, for example, for $a_0$ and $a_2$
\begin{equation}\label{a0a2}
a_0=\frac{\langle\omega_1^0\,h_{17}\rangle}{\sqrt{2\pi}|\sigma|}, \qquad a_2=\frac{\langle\omega_1^2\,h_{17}\rangle-\sigma^2\langle\omega_1^0\,h_{17}\rangle}{4\sqrt{2\pi}|\sigma|^3}.
\end{equation}

To further constrain $h_{17}(\omega_1,\mu)$, we use the fact that it is a soft function. We limit its absolute value to $1$ GeV, i.e. $|h_{17}(\omega_1,\mu)|\leq$ 1 GeV and require, as in \cite{Benzke:2010js}, that it should not have any significant structures, such as peaks or zeros, outside the range $|\omega_1|\leq 1$ GeV. This allows us to restrict the range of $\sigma$. For example, assuming a model of a sum of two Hermite polynomials, for given values of $\langle\omega_1^0\,h_{17}\rangle$ and $\langle\omega_1^2\,h_{17}\rangle$, the requirement on significant structures only for $|\omega_1|\leq 1$ GeV gives an upper bound on $\sigma$ and the condition $|h_{17}(\omega_1,\mu)|\leq$ 1 GeV gives a lower bound on $\sigma$. For example, assuming the central values for  $\langle\omega_1^0\,h_{17}\rangle=0.237 \mbox{ GeV}^2$ and $\langle\omega_1^2\,h_{17}\rangle=0.15\mbox{ GeV}^4$ gives $0.27\mbox{ GeV}<\sigma<0.62\mbox{ GeV}$. For other values of $\langle\omega_1^0\,h_{17}\rangle$ and   $\langle\omega_1^2\,h_{17}\rangle$ within their one standard deviation range, the range of $\sigma$ can be larger, but we restrict $\sigma$ to be less than 1 GeV. As we will see below, this does not affect our estimates in practice since the extremal values we obtain are for $\sigma<$ 1 GeV anyway. 

We consider models with one, and two Hermite polynomials whose coefficients are determined by the known moments as well as models with more Hermite polynomials whose coefficients depended on unknown moments.  

\subsubsection{One Hermite polynomial model}
Since $\sigma$ is not determined by the moments, a model with one Hermite polynomial can in principle be adjusted to fit both the zero and second moment of $h_{17}$. Notice from (\ref{a0a2}) and (\ref{moments_numerical}) that $a_0$ is never zero, so a one Hermite polynomial model must include $H_0$.  The one Hermite polynomial model is thus
\begin{equation}
h^{\mbox{\scriptsize model-1}}_{17}(\omega_1)=\frac{\langle\omega_1^0\,h_{17}\rangle}{\sqrt{2\pi}|\sigma|}e^{-\frac{\omega_1^2}{2\sigma^2}}.
\end{equation}
The second moment of $h^{\mbox{\scriptsize model-1}}_{17}$ implies $\sigma=\sqrt{\langle\omega_1^2\,h_{17}\rangle/\langle\omega_1^0\,h_{17}\rangle}$. This is also the condition for $a_2=0$ in (\ref{a0a2}). 

Varying the zero and second moment within their one standard deviation ranges leads to values of $\sigma$ that exceed 1 GeV. For example, this happens for  $\langle\omega_1^2\,h_{17}\rangle=0.27\mbox{ GeV}^4$ and for almost all values of $\langle\omega_1^0\,h_{17}\rangle$ within its one standard deviation range. Based on the criterion above, we should reject such models. But even if we include them, the values of $\Lambda_{17},\tilde\Lambda_{17}^u$, and $\tilde\Lambda_{17}^c$ we obtain are included in the ranges for the two Hermite polynomials model below. Thus we find $\Lambda_{17}\in[-8,-1]$ MeV, $\tilde\Lambda_{17}^c\in[0,7.5]$ MeV, and $\tilde\Lambda_{17}^u\in[45,220]$ MeV.

\subsubsection{Sum of two Hermite polynomials model}\label{subsec:2H} 
A model that contains a sum of two Hermite polynomials for a given value of $\sigma$ is determined by (\ref{a0a2}), i.e. by the zeroth and second moment of $h_{17}$. The values of $a_0$ and $a_2$ depend on $\sigma$, but for $\sigma$ of the order of a few hundred MeVs they are typically of the order of 1 GeV and often smaller. 

Numerically scanning over the one standard deviation range of the moments and  the possible values of $\sigma$ in increments of  $\delta\sigma=0.01$ GeV, and based on the restrictions above on $h_{17}$ gives $\Lambda_{17}\in[-21,-1]$~MeV. The lower value is obtained for  $\langle\omega_1^0\,h_{17}\rangle=0.197\mbox{ GeV}^2$, $\langle\omega_1^2\,h_{17}\rangle=0.27\mbox{ GeV}^4$, $\sigma=0.44$ GeV, $m_c=1.17$ GeV, and $m_b=4.61$ GeV. The upper value is obtained for $\langle\omega_1^0\,h_{17}\rangle=0.277\mbox{ GeV}^2$, $\langle\omega_1^2\,h_{17}\rangle=0.03\mbox{ GeV}^4$, $\sigma=0.14$ GeV, $m_c=1.23$ GeV, and $m_b=4.55$ GeV. Thus the extremal values are obtained for extremal values of the two moments, anti-correlated, and the extremal values of $m_c$ and $m_b$, anti-correlated.

It is instructive to check the dependance on $m_b$ and $m_c$. For example, consider the set $\langle\omega_1^0\,h_{17}\rangle=0.197\mbox{ GeV}^2$, $\langle\omega_1^2\,h_{17}\rangle=0.27\mbox{ GeV}^4$, $\sigma=0.44$ GeV that leads to $\Lambda_{17}=-21$ MeV. Changing $m_b=4.61$ to $m_b=4.55$ GeV  while keeping $m_c=1.17$ GeV changes $\Lambda_{17}$ by $+1$ MeV. Thus the dependance on the value of $m_b$ is rather mild.  Changing $m_c=1.17$ GeV to $m_c=1.23$ GeV while keeping $m_b=4.61$ GeV changes $\Lambda_{17}$ by $+6$ MeV. Thus the dependance on the value of $m_c$ is more pronounced. We will further comment on this in the conclusions.  

Using the same method we can find the range of allowed values for $\tilde\Lambda_{17}^c$. We have $\tilde\Lambda_{17}^c\in[0,10]$~MeV. The lower value is obtained for  $\langle\omega_1^0\,h_{17}\rangle=0.277\mbox{ GeV}^2$, $\langle\omega_1^2\,h_{17}\rangle=0.03\mbox{ GeV}^4$, $\sigma=0.14$ GeV, $m_c=1.13$ GeV, and $m_b=4.55$ GeV. The upper value is obtained for $\langle\omega_1^0\,h_{17}\rangle=0.197\mbox{ GeV}^2$, $\langle\omega_1^2\,h_{17}\rangle=0.27\mbox{ GeV}^4$, $\sigma=0.58$ GeV, $m_c=1.07$ GeV, and $m_b=4.61$ GeV. Again the extremal values are obtained for extremal values of the two moments, anti-correlated, and the extremal values of $m_c$ and $m_b$, anti-correlated.

Lastly, we consider $\tilde\Lambda_{17}^u$. Using the parameterization above we have the expression 
\begin{equation} 
\tilde\Lambda_{17}^u = \frac23\,h_{17}(0)=\dfrac{3\sigma^2\langle\omega_1^0\,h_{17}\rangle-\langle\omega_1^2\,h_{17}\rangle}{3\sqrt{2\pi}|\sigma|^3}.
 \end{equation}
Since both moments are positive within their one standard deviation range, we can easily make $h_{17}(0)$ negative by choosing a small value of $\sigma$. Thus the smallest value of  $h_{17}(0)$ based on  $|h_{17}(\omega_1,\mu)|\leq1$ GeV is $-1$ GeV. For example, for the central values of $\langle\omega_1^0\,h_{17}\rangle$ and $\langle\omega_1^2\,h_{17}\rangle$, the value of $\sigma=0.27$ GeV gives $h_{17}(0)=-1$ GeV. To make $h_{17}(0)$ reach its highest possible value, we can choose the smallest value of $\langle\omega_1^2\,h_{17}\rangle$, 0.03 GeV$^4$ and the largest value of  $\langle\omega_1^0\,h_{17}\rangle$, 0.277 GeV$^2$. The extremal value of  $h_{17}(0)=0.33$ GeV is obtained for $\sigma =\sqrt{\langle\omega_1^2\,h_{17}\rangle/\langle\omega_1^0\,h_{17}\rangle}=0.33$ GeV. Based on this we find that $\tilde\Lambda_{17}^u\in[-660,220]$ MeV.  

\subsubsection{Sum of three Hermite polynomials model} 
A model that contains a sum of three Hermite polynomials for a given value of $\sigma$  requires the knowledge of the fourth moment of $h_{17}$:
\begin{equation}\label{a4}
a_4=\frac{\langle\omega_1^4\,h_{17}\rangle-6\sigma^2\langle\omega_1^2\,h_{17}\rangle+3\sigma^4\langle\omega_1^0\,h_{17}\rangle}{96\sqrt{2\pi}|\sigma|^5}.
\end{equation}
The fourth moment is currently unknown since it is a matrix element of a dimension 9 operator. To asses the impact of such a moment if it were known, we assume the very conservative estimate of $[-0.3,0.3]$ GeV$^6$ for $\langle\omega_1^4\,h_{17}\rangle$. This covers all the numerical ranges in (\ref{moments_numerical}) but with a different dimension of course. We still maintain the restrictions of the values, zeros, and extremal points of $h_{17}$ to be below 1 GeV. 

Numerically scanning over the one standard deviation range of the known zero and second moments, the range $[-0.3,0.3]$ GeV$^6$ for the unknown fourth moment in increments of 0.05 GeV and  the possible values of $\sigma$ based on the restrictions above gives $\Lambda_{17}\in[-24,3]$~MeV. The lower value is obtained for  $\langle\omega_1^0\,h_{17}\rangle=0.277\mbox{ GeV}^2$, $\langle\omega_1^2\,h_{17}\rangle=0.27\mbox{ GeV}^4$, $\langle\omega_1^4\,h_{17}\rangle=0.3\mbox{ GeV}^6$, $\sigma=0.32$ GeV, $m_c=1.17$ GeV, and $m_b=4.61$ GeV. The upper value is obtained for $\langle\omega_1^0\,h_{17}\rangle=0.237\mbox{ GeV}^2$, $\langle\omega_1^2\,h_{17}\rangle=0.03\mbox{ GeV}^4$, $\langle\omega_1^4\,h_{17}\rangle=-0.1\mbox{ GeV}^6$, $\sigma=0.34$ GeV, $m_c=1.17$ GeV, and $m_b=4.61$ GeV. The obtained range is only slightly different from the two Hermite polynomial model and reflects our generous range for the unknown fourth moment. 

Similarly we find the range for $\tilde\Lambda_{17}^c$. The positive values are included in the range obtained for a sum of two Hermite polynomials. We also get negative values in the range $[-5.6,0]$ MeV. The smallest value is obtained for $\langle\omega_1^0\,h_{17}\rangle=0.277\mbox{ GeV}^2$, $\langle\omega_1^2\,h_{17}\rangle=0.03\mbox{ GeV}^4$, $\langle\omega_1^4\,h_{17}\rangle=-0.11\mbox{ GeV}^6$, $\sigma=0.34$ GeV, $m_c=1.07$ GeV, and $m_b=4.61$ GeV.

Unlike the two Hermite polynomial model we can make $h_{17}(0)$ reach a value of  $1$ GeV.  For example, taking the central values of the zeroth and second moment $\langle\omega_1^0\,h_{17}\rangle=0.237\mbox{ GeV}^2$, $\langle\omega_1^2\,h_{17}\rangle=0.15\mbox{ GeV}^4$ we find that for $\langle\omega_1^4\,h_{17}\rangle=0.1\mbox{ GeV}^6$ and $\sigma=0.25$ GeV $h_{17}(0)=1$ GeV. This result is not surprising. The moments are global properties of the function and it is hard to restrict using them values of the function at a single point. We conclude that for this model $\tilde\Lambda_{17}^u$ can be as large as 660 MeV, which is the largest value possible under the condition $|h_{17}(\omega_1,\mu)|\leq1$ GeV. 

\subsubsection{Sum of four Hermite polynomials model} 
To test how typical is the change from a model with two Hermite polynomials to a model with three Hermite polynomials, we consider a model with four Hermite polynomial. We assume again the very conservative estimate of $[-0.3,0.3]$ GeV$^8$ for the sixth moment $\langle\omega_1^6\,h_{17}\rangle$ that determines the coefficient of $H_6$ in (\ref{h17General}).  As in the three Hermite polynomials model we assume the range of $[-0.3,0.3]$ GeV$^6$ for $\langle\omega_1^4\,h_{17}\rangle$. 

Scanning over the values of the fourth and sixth moment we find that the smallest value of $\Lambda_{17}$ is $-22$ MeV, i.e. in the range we obtained for three Hermite polynomials. The highest value we obtain is  $5$~MeV for $\langle\omega_1^0\,h_{17}\rangle=0.277\mbox{ GeV}^2$, $\langle\omega_1^2\,h_{17}\rangle=0.03\mbox{ GeV}^4$, $\langle\omega_1^4\,h_{17}\rangle=-0.1\mbox{ GeV}^6$, $\langle\omega_1^6\,h_{17}\rangle=-0.2\mbox{ GeV}^8$, $\sigma=0.29$ GeV, $m_c=1.17$ GeV, and $m_b=4.61$ GeV. This should be compared to the maximum value of $-1$ MeV and $3$ MeV for the two and three Hermite polynomial models, respectively.   

For $\tilde\Lambda_{17}^c$ we find positive values that are already included in the ranges of the two and three Hermite polynomial models above. The smallest negative value we find for $\tilde\Lambda_{17}^c$ is $-7$ MeV for $\langle\omega_1^0\,h_{17}\rangle=0.277\mbox{ GeV}^2$, $\langle\omega_1^2\,h_{17}\rangle=0.03\mbox{ GeV}^4$, $\langle\omega_1^4\,h_{17}\rangle=-0.1\mbox{ GeV}^6$, $\langle\omega_1^6\,h_{17}\rangle=-0.2\mbox{ GeV}^8$, $\sigma=0.29$ GeV, $m_c=1.07$ GeV, and $m_b=4.61$ GeV.

Since $\tilde\Lambda_{17}^u$ obtains its smallest and largest possible values for the two and three Hermite polynomial models, there is no need to check the effect of the four Hermite polynomials model.

\subsubsection{Sum of five and six Hermite polynomials model} 
One can consider continuing in this way and check models with five and even six Hermite polynomials, based on similar assumptions for the unknown moments. Namely, we assume that the unknown $k$-th moment is in the range $[-0.3,0.3]$ GeV$^{\,k+2}$. Scanning over the ranges in increments of $0.1$ GeV$^{\,k+2}$ we find that there are no solutions that satisfy our requirements on $h_{17}(0)$. One reason is the fast growth of the value of $H_n(0)$. To maintain a value of $|h_{17}(0)|\leq 1$ GeV requires that the coefficient of $H_n(0)$ be increasingly smaller.

\subsubsection{Summary}\label{subsec:summary} 
Using a two Hermite polynomial model we find $\Lambda_{17}\in[-21,-1]$~MeV, $\tilde\Lambda_{17}^c\in[0,10]$~MeV, and $\tilde\Lambda_{17}^u\in[-660,220]$ MeV. Using a three Hermite polynomial model and \emph{assuming} the range $[-0.3,0.3]$ GeV$^6$ for $\langle\omega_1^4\,h_{17}\rangle$ we find $\Lambda_{17}\in[-24,3]$~MeV. The range for $\tilde\Lambda_{17}^c$ can include values $\in[-5.6,0]$~MeV. Also, $\tilde\Lambda_{17}^u$ can be as large $660$ MeV, which is it the largest possible value based on our assumptions for $h_{17}$. Using the four Hermite polynomial model with similar assumptions on the fourth and sixth moments changes the highest value of $\Lambda_{17}$ to 5 MeV and the lowest value of $\tilde\Lambda_{17}^c$ to $-7$ MeV.

Combining the results above and rounding to the closest integer we have $\Lambda_{17}\in[-24,5]$~MeV, $\tilde\Lambda_{17}^c\in[-7,10]$~MeV, and $\tilde\Lambda_{17}^u\in[-660,660]$ MeV.  

\subsection{Phenomenological estimates}\label{subsec:pheno}
Based on the analysis above we can update the results of \cite{Benzke:2010js} and \cite{Benzke:2010tq}. To highlight the changes in the uncertainties, we follow the same methodology in these papers to obtain the uncertainties on the total rate and the CP asymmetry. 

For the total rate we use $\Lambda_{17}\in[-24,5]$~MeV, equation (\ref{F17def}), and the values $C_1(\mu) = 1.257$, $C_{7\gamma}(\mu) = -0.407$ (calculated at $\mu=1.5$ GeV), and $m_b=4.58$ GeV to obtain 
\begin{equation}
{\cal F}_E\big|_{17}\in[-0.3,+1.6]\,\%.   
\end{equation}
This should be compared  to the range $[-1.7,+4.0]\,\%$ in \cite{Benzke:2010js}. 

To obtain the total uncertainty we use ${\cal F}_E\big|_{88}\in[-0.3,+1.9]\,\%$ from  \cite{Benzke:2010js}. For  ${\cal F}_E\big|_{78}$, we can use either the VIA value from  \cite{Benzke:2010js} ${\cal F}_E\big|_{78}^{\rm VIA}\in[-2.8, -0.3]\,\%$ or the new experimental value discussed in the introduction  ${\cal F}_E\big|_{78}^{\rm exp} \in [-1.4, +2]\,\%$. Scanning over the various contributions gives 
\begin{equation}\label{NewFVIA}
   -3.4\% < {\cal F}_E(\Delta) < +3.2\%
   \quad \mbox{(using VIA)} \,,
\end{equation}
using the theoretical estimate for ${\cal F}_E\big|_{78}^{\rm VIA}$. Compared to $-4.8\% < {\cal F}_E(\Delta) < +5.6\% \,\mbox{(using VIA)}$ in \cite{Benzke:2010js},  the new estimate reduces the total error by about a third. Using the experimental estimate instead, the range is 
\begin{equation}\label{NewFExp}
   -2.0\% < {\cal F}_E(\Delta) < +5.5\%
   \quad \mbox{(using $\Delta_{0-}$)} \,.
\end{equation}
Compared to  $-6.4\% < {\cal F}_E(\Delta) < +11.5\%  \,\mbox{(using } \Delta_{0-})$ in \cite{Benzke:2010js}, the new estimate reduces the total error by about a half. 

For the CP asymmetry we use $\tilde\Lambda_{17}^c\in[-7,10]$~MeV and $\tilde\Lambda_{17}^u\in[-660,660]$ MeV. The expressions in \cite{Benzke:2010tq} can be used to find the resulting CP asymmetry. For example, the sum of the direct and resolved contributions to the CP asymmetry in the SM is  \cite{Benzke:2010tq}
\begin{equation}\label{ACPSM}
 {\cal A}_{X_s\gamma}^{\rm SM}= \left(1.15\times\frac{\tilde\Lambda_{17}^u-\tilde\Lambda_{17}^c}{300\,\mbox{MeV}}
    + 0.71 \right) \% \,,
\end{equation}
where we use the same\footnote{Using the values  $C_1(\mu) = 1.204$, $C_{7\gamma}(\mu) = -0.378$, $m_c=1.1$ GeV (calculated at $\mu=2.0$ GeV), and $m_b=4.58$ GeV changes 1.15 to 1.18 and 0.71 to 0.70 in (\ref{ACPSM}). Because of rounding there is not change to the allowed range for  ${\cal A}_{X_s\gamma}^{\rm SM}$.} parameters as in \cite{Benzke:2010tq}. Because of the increase in the allowed range for $\tilde\Lambda_{17}^u$,  the allowed range for ${\cal A}_{X_s\gamma}^{\rm SM}$ increases to $-1.9\%<{\cal A}_{X_s\gamma}^{\rm SM}<3.3\%$ . This should be compared to the range $-0.6\%<{\cal A}_{X_s\gamma}^{\rm SM}<2.8\%$ in \cite{Benzke:2010tq}. Similar updates can be applied to the resolved $Q_1^q-Q_{7\gamma}$ contributions to the CP asymmetry in extensions of the SM.

\section{Conclusions and outlook} \label{sec:conclusions}
Resolved photon contributions limit the theoretical uncertainty achievable in  $\bar B\to X_s\gamma$ decay. They give the dominant uncertainty ($\sim5\%$) on the SM prediction of the total rate and give the dominant effect within the SM for the CP asymmetry. In this paper we have used recent progress in the knowledge of HQET parameters to reevaluate resolved photon contribution from the interference of $Q_1^q$ ($q=u,c$) and $Q_{7\gamma}$. 

This contribution depends on a soft function $g_{17}(\omega,\omega_1,\mu)$, defined in (\ref{g17def}), whose moments are related to HQET parameters.  In section \ref{sec:moments} we presented such relations for general moments in $\omega$ alone, see (\ref{Generalw}), in $\omega_1$ alone, see (\ref{Generalw1}), and general mixed moments in  $\omega$ and $\omega_1$, see (\ref{Generalww1}). The general decomposition of HQET matrix elements presented in \cite{Gunawardana:2017zix} allows to relate these matrix elements to the basis \cite{Mannel:2010wj} of HQET parameters arising from operators of up to (and including) dimension 8 used inclusive semileptonic $B$ decays.  

In section \ref{sec:applications} we presented several phenomenological applications for the moment relations. First, numerical values of HQET parameters arising from operators of up to (and including) dimension 8  were obtained in \cite{Gambino:2016jkc} from a global fit to semileptonic $B$ decay spectra and information based on the lowest-lying state saturation approximation in \cite{Mannel:2010wj,Heinonen:2014dxa}. These allow to find numerical ranges for eight non-zero lowest moments of $g_{17}(\omega,\omega_1,\mu)$. Second, \emph{assuming} a limited support for $h_{17}(\omega_1)$, defined in (\ref{h17def}), we have used the values of the moments to estimate the $Q_1^c-Q_{7\gamma}$ contribution to the total rate. We have also investigated the convergence of the expansion of the penguin function in $m_b\,\omega_1/m_c^2$ and the resolved photon contribution in $\Lambda_{\mbox{\scriptsize QCD}}/m_b$. The former converges slowly, consistent with the power counting of $m_c^2\sim m_b\Lambda_{\mbox{\scriptsize QCD}}$. The latter exhibits the expected power suppression with the exception of the first  term which is accidentally small due to $h_{17}$ being an even function.  Third, since the assumption about the support of $h_{17}$ is known to be too restrictive  \cite{Benzke:2010js}, we considered a systematic improvable model for $h_{17}$. 

To construct the model we use properties of  $h_{17}(\omega_1)$.  It is an even function of $\omega_1$. While its argument formally takes values for $-\infty\leq\omega_1\leq\infty$, it is a soft function, so one expects it not to have significant structures beyond $|\omega_1|\leq1$ GeV. Similarly we limit its values to $|h_{17}(\omega_1)| \leq1$ GeV. The model we use is to express $h_{17}(\omega_1)$ as sum of even Hermite polynomials multiplied by a Gaussian of width $\sigma$, see (\ref{h17General}). The coefficients of the polynomials are determined by the moments of $h_{17}(\omega_1)$. A benefit of such an approach is that the lack of knowledge of higher moments does not affect models that only use lower moments.  The value of $\sigma$ can be constrained by the requirement on the support and values of $h_{17}(\omega_1)$. 

The current numerical knowledge of the moments allows us to determine the coefficients of a sum of the first two even Hermite polynomials. Using the requirements above, we numerically scan over the values of $\sigma$ to determine the extremal possible values for the parameters $\Lambda_{17},\tilde\Lambda_{17}^c,\tilde\Lambda_{17}^u$, defined in section \ref{subsec:resolved}. We also consider models with higher number of even Hermite polynomials by assuming a conservative range for the unknown moments.  Taking the envelopes of our results, we find the estimates for the parameters in section  \ref{subsec:summary}. For $\Lambda_{17}$ and $\tilde\Lambda_{17}^c$ the new analysis finds a reduction in the allowed range compared to \cite{Benzke:2010js, Benzke:2010tq}. For  $\tilde\Lambda_{17}^u$ we find an increase in the allowed range compared to \cite{Benzke:2010tq} since it depends on the value of $h_{17}(0)$ which is not well constrained by global properties like moments. In section \ref{subsec:pheno} we give estimates for the total rate uncertainty and the SM prediction for the CP asymmetry from resolved photon effects from $Q_1^q-Q_{7\gamma}$.  For the total rate the uncertainty is reduced, but for the CP asymmetry the uncertainty is increased.   

We conclude with a discussion of possible future improvements. With the new moment information we get a better handle on hadronic effects. As a result,   uncertainties from perturbative effects become more conspicuous. For example, following \cite{Benzke:2010js, Benzke:2010tq} the scale $\mu$ for the Wilson coefficients and the charm quark mass was taken to be 1.5 GeV for the total rate and 2 GeV for the CP asymmetry. Because the resolved photon contribution are currently treated at leading order in $\alpha_s$, the scale dependance is not controlled. In order to improve on that, one needs to calculate $\alpha_s$ corrections to the resolved $Q_1^c-Q_{7\gamma}$ contribution.  Controlling the scale dependance can also help to better estimate the uncertainty from the value of the charm quark mass. See section \ref{subsec:2H} for an example of the numerical effect of the charm quark mass. 

The modeling can improve considerably if the numerical value of matrix elements of dimension 9 HQET operators were known. The first step of classifying the possible spin-dependent dimension 9 operators can be easily carried out using the methods of \cite{Gunawardana:2017zix}. With the expected Belle II data, it is conceivable that the work of \cite{Mannel:2010wj} can be extended to this level and a similar analysis to \cite{Gambino:2016jkc} can be applied to the data. 

Finally, we have considered quantities  that are integrated over the photon energy. One can consider also the photon energy spectrum itself, or its moments. The moment information above can be used to model the $Q_1^q-Q_{7\gamma}$ resolved photon contribution in this case. This is left for a future work.

\vskip 0.2in
\noindent
{\bf Acknowledgements}
\vskip 0.1in
\noindent
We thank Paulo Gambino, Mikolaj Misiak, and Claude Pruneau for useful discussions. G.P. thanks the Technion and the Weizmann Institute for their hospitality and support during the completion of this work. This work was supported by the U.S. Department of Energy grant DE-SC0007983, by a grant from the Simons Foundation (562836, G.P.), and by a fellowship from the Lady Davis Foundation at the Technion.

\begin{appendix}

\section{Appendix: A useful identity }\label{appendix}
The  Wilson line 
\begin{equation}
S_{\bar n}(x)=\mbox{\bf P}\exp \left( ig\int_{-\infty}^0 \,du \,\bar n\cdot A_s(x+u\bar n) \right)\,,
\end{equation}
obeys the equation $i\bar n\cdot D\,S_{\bar n}(x)=0$, where $iD^\mu=i\partial^\mu+gA^\mu$, see, e.g., \cite{Becher:2014oda} for a derivation.  Thus $i\bar n\cdot\partial S_{\bar n}(x)=-g\bar n\cdot A(x)S_{\bar n}(x)$. Taking the Hermitian conjugate of this identity gives $i\bar n\cdot\partial S^\dagger_{\bar n}(x)=S^\dagger_{\bar n}(x)g\bar n\cdot A(x)$. Consider now $i\bar n\cdot \partial \left(S_{\bar n}^\dagger(x) O(x) S_{\bar n}(x) \right)$, where $O(x)$ is an operator. Using the identities above we have 
\begin{eqnarray}
&&i\bar n\cdot \partial \left(S_{\bar n}^\dagger(x) O(x) S_{\bar n}(x) \right)=\nonumber\\
&=&\big(i\bar n\cdot \partial S_{\bar n}^\dagger(x)\big)O(x) S_{\bar n}(x) +S_{\bar n}^\dagger(x)\big(i\bar n\cdot \partial\, O(x)\big)S_{\bar n}(x) +S_{\bar n}^\dagger(x)O(x) \big(i\bar n\cdot \partial S_{\bar n}(x) \big)\nonumber\\
&=&S^\dagger_{\bar n}(x)g\bar n\cdot A(x)O(x) S_{\bar n}(x) +S_{\bar n}^\dagger(x)\big(i\bar n\cdot \partial\, O(x)\big)S_{\bar n}(x)-S_{\bar n}^\dagger(x) O(x) g\bar n\cdot A(x)S_{\bar n}(x)=\nonumber\\
&=& S^\dagger_{\bar n}(x)[g\bar n\cdot A(x),O(x)] S_{\bar n}(x)+S_{\bar n}^\dagger(x)\big[i\bar n\cdot \partial ,O(x)\big]S_{\bar n}(x)=S_{\bar n}^\dagger(x)\big[i\bar n\cdot D,O(x)\big]S_{\bar n}(x).
\end{eqnarray}
In the last line we have used the identity $\big[i\bar n\cdot \partial ,O(x)\big]f(x)=\big(i\bar n\cdot \partial\, O(x)\big)f(x)$ for an arbitrary function $f(x)$. Thus we have the identity 
 \begin{equation}
 i\bar n\cdot \partial \left(S_{\bar n}^\dagger(x) O(x) S_{\bar n}(x) \right)=S_{\bar n}^\dagger(x)\big[i\bar n\cdot D,O(x)\big]S_{\bar n}(x).
 \end{equation}
\end{appendix}

\end{document}